\documentclass[aps,prd,twocolumn,a4paper]{revtex4-1}
\usepackage{ulem}
\usepackage{color}
\usepackage{graphicx}
\usepackage{hyperref}   
\usepackage{appendix}
\usepackage{epsfig}
\usepackage{epstopdf}
\usepackage{amsmath}
\usepackage{subfig}
\usepackage{comment}
\usepackage[dvipsnames]{xcolor}
\newcommand{\be}{\begin{equation}}
\newcommand{\ee}{\end{equation}}
\newcommand{\ba}{\begin{eqnarray}}
\newcommand{\ea}{\end{eqnarray}}

\begin{document}
\title{Bulk viscosity of a hot QCD/QGP
medium in strong magnetic field within relaxation-time approximation}
\author{Manu Kurian}
\email{manu.kurian@iitgn.ac.in}
\author{Vinod Chandra }
\email{vchandra@iitgn.ac.in}
\affiliation{Indian Institute of Technology 
Gandhinagar, Gandhinagar-382355, Gujarat, India}

\begin{abstract}
The bulk viscosity of hot QCD medium has been obtained in the 
presence  of strong magnetic field.  The present investigation involves 
the estimation of the quark damping rate and subsequently 
the thermal relaxation time for quarks in the presence of magnetic field while 
realizing the hot QCD  medium as an effective Grand-canonical ensemble of 
effective gluons and quarks-antiquarks. The dominant process in 
the strong field limit is $1\rightarrow 2$ ($g\rightarrow q \bar{q}$) 
which contributes to the bulk viscosity in a most significant way.  Further, setting up the linearized transport equation in the framework 
of an effective kinetic theory with hot QCD medium effects 
and employing the relaxation time approximation, the bulk viscosity 
has been estimated in lowest Landau level  (LLL) and beyond.  
The temperature dependence of the ratio of the 
 bulk viscosity to entropy density indicates towards its 
 rising behavior  near the transition temperature.
\\

\end{abstract}

\maketitle
 
\section{Introduction}
Relativistic heavy-ion collision   experiments (RHIC)  set the platform for the 
creation and study of quark-gluon plasma (QGP) as a
near-perfect fluid~\cite{STAR,Aamodt:2010pb}. Recent investigations on the QGP 
suggests the presence of 
extremely high magnetic field  in the 
early stages of the collisions, (specially in the 
non-central asymmetric collisions)~\cite{Skokov:2009qp,Zhong:2014cda,deng,Das:2016cwd}. 
In this context, a  deeper understating of various aspects of the  QGP in 
the strong magnetic field is the prime focus of the current research on the physics of the RHIC.
In particular,  Chiral magnetic effect~\cite{Fukushima:2008xe,Sadofyev:2010pr, Huang} 
and chiral vortical effects~\cite{Kharzeev:2015znc,Avkhadiev:2017fxj,Yamamoto:2017uul} 
gained huge attention in the QGP community. More recently, the discovery of global $\Lambda$-hyperon polarization 
in non-central RHIC~\cite{STAR:2017ckg,Becattini:2016gvu} 
opens up  a new direction in the study of  the QGP in the presence of strong magnetic field.

 Recall that the  quark-antiquark pair production and fusion processes are kinematically
possible in the presence of the strong magnetic 
field~\cite{Fukushima:2011nu, Tuchin:2010gx} via $1\rightarrow 2$ processes that
dominate  over $2\rightarrow 2$ scattering processes while  estimating the  transport
coefficients. This could be understood in terms of the fact that the  rate is proportional to coupling constant $\alpha_{s}$ in the 
case of the former, whereas that of the binary processes, it  is proportional to $\alpha_{s}^{2}$~\cite{Hattori:2016cnt}.
The magnetic field effects enter in the quark-antiquark degrees of freedom through the Landau levels. The  strong magnetic field restricts the 
calculation to the (1+1)- dimensional ground state  i.e., lowest Landau level (LLL)~\cite{Bhattacharya:2007vz,
Bruckmann:2017pft} (the  dimensional  reduction). On the other hand, the electrically chargeless gluons are not directly coupled to the 
magnetic field through dispersion relation. However, the
gluonic dynamics in the presence of magnetic field can be affected through the 
quark loop while defining the gluon vertex through the self-energy where the quark/antiquark loop contributes.

The quantitative study of the transport coefficients in the
hot QCD medium is required for the estimation of the experimental
observables like transverse momentum spectra and collective flow of  the QGP within the
dissipative relativistic hydrodynamic framework. 
In particular, extremely low viscosity to entropy ratio indicate 
the larger elliptic flow observed in RHIC. Besides providing the basis for 
understanding the probes of QGP, the transport coefficients give 
insights to the electromagnetic response of the medium. 
Recently, a number of ALICE results have shown the relevance of transport processes in 
the RHIC~\cite{,Adam:2016izf,Abelev:2012pa,Abelev:2012pp}. 
Since the strong magnetic field is generated in the non-central asymmetric HIC, 
the dissipative magnetohydrodynamics describes the 
transport process of the medium. This  sets the strong motivation 
for the estimation of transport coefficients of the QGP 
in presence of the strong magnetic field.

There have been several attempts to estimate the transport 
coefficients of the hot QCD medium 
in the strong magnetic field~\cite{Hattori:2016lqx,
Hattori:2017qih,Yin:2013kya,Critelli:2014kra,Li:2017tgi}. In a very recent work, Fukushima and  Hidaka~\cite{Fukushima:2017lvb},  estimated
 the longitudinal conductivity in the magnetic field  beyond LLL approximation by solving the kinetic equation, considering the scattering amplitude of synchrotron radiation and the pair annihilation processes.  The authors have numerically 
 shown that the contribution from LLL is the dominant one.
 
 The goal 
of the present investigations is to estimate the 
temperature dependence of the thermal relaxation time and thereby the  
effective bulk viscosity while  encoding the hot QCD medium effects in strong 
magnetic field background through an effective quasi-particle model. The analysis has 
been done with relativistic semi-classical transport theory, 
in which, microscopic particle interactions bridges to macroscopic 
transport phenomena of the thermodynamic system. 
The kinetic theory approach is followed within the linear 
response analysis of transport equation in which
magnetic field enters through the propagator (matrix element in
collision integral) and momentum distribution functions of the quarks and antiquarks. 
Note that another equivalent  approach to investigate the transport coefficients of the hot
QCD in the magnetic field background is the hard thermal loop 
effective theory (HTL)~\cite{Rath:2017fdv,Bandyopadhyay:2017cle}. 
We are following the former one
here.

Hot QCD medium effects encrypted as the equation of sate (EoS) 
dependence on the transport coefficients within effective linear transport theory are 
well understood~\cite{Chandra:2008hi,Mitra:2016zdw,Mitra:2018akk, 
Bluhm:2011,Chandra:2012qq, Kapusta:2016, Das:2012ck}. 
In~\cite{Kurian:2017yxj}, the authors have recently
estimated the EoS/medium  dependence on the longitudinal electrical 
conductivity for the $1\rightarrow 2$ processes in the strong magnetic field background. 
In the present work, we followed the effective fugacity 
quasiparticle model (EQPM), proposed in~\cite{Chandra:2011en, Chandra:2007ca} 
and extended in the case of the strong
magnetic in Ref~\cite{Kurian:2017yxj}. The  first step 
towards the evaluation the bulk viscosity is the 
quark damping rate $\Gamma_{eff}$ in the strong field limit that leads to the 
thermal relaxation time $\tau_{eff}$, followed by the 
estimation of the bulk viscosity $\zeta_{eff}$ 
in the presence of magnetic field by setting up an 
effective linearized transport equation.
This has been done not only in LLL but also with the  higher Landau level 
(HLL) corrections. 
 
 The paper is organized as follows. Section II deals with the 
mathematical formalism for the estimation 
of the effective thermal relaxation time 
and the  bulk viscosity along with the description of 
hot QCD effective coupling constant with HLL corrections. 
Section III constitutes the predictions on the bulk 
viscosity and the related discussions. Finally, 
in section IV, the conclusion and outlook of the work are presented.

\begin{figure*}
 \centering
 \subfloat{\includegraphics[height=7.6cm,width=7.705cm]{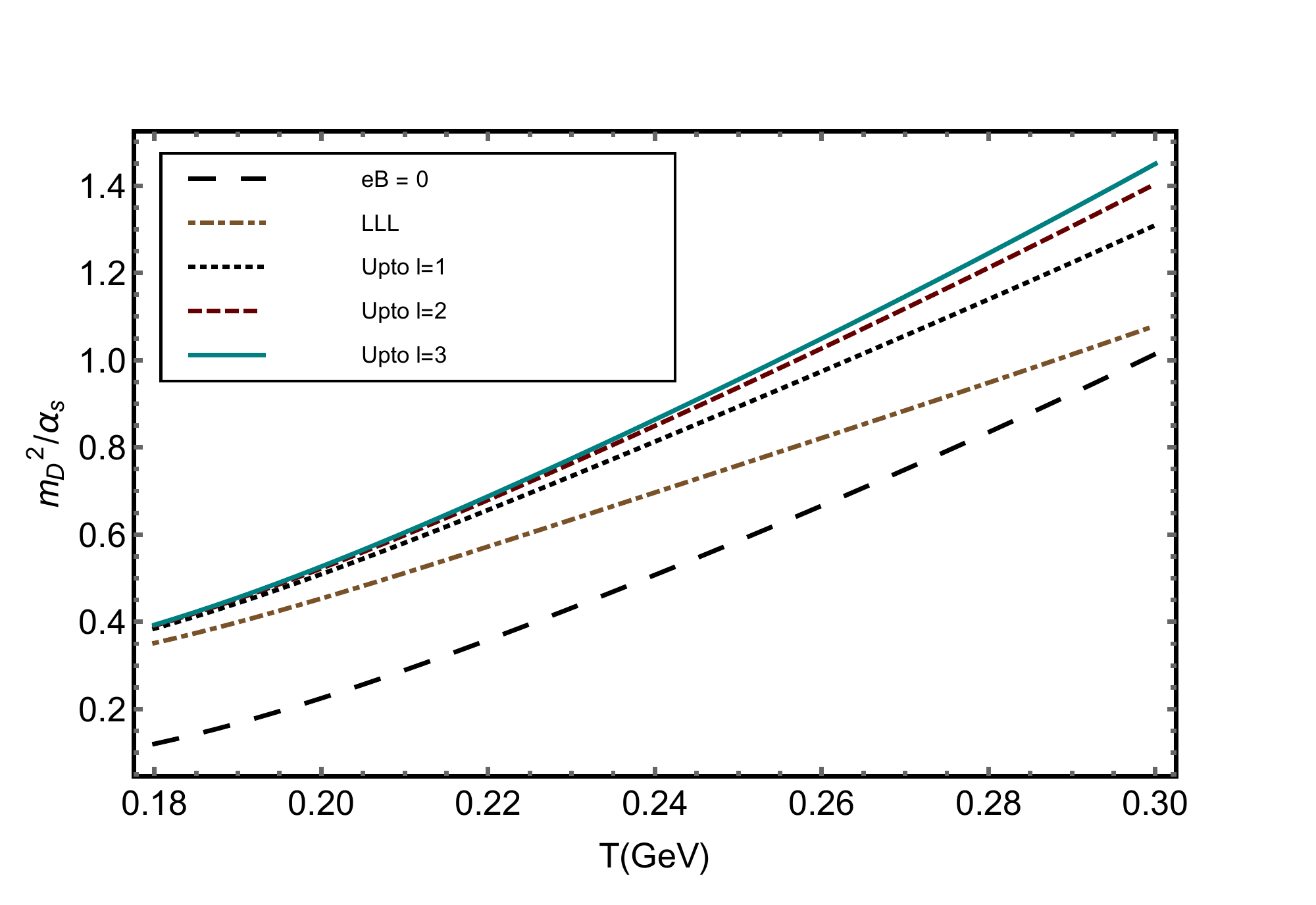}}
 \hspace{1 cm}
 \subfloat{\includegraphics[height=6.80cm,width=8.80cm]{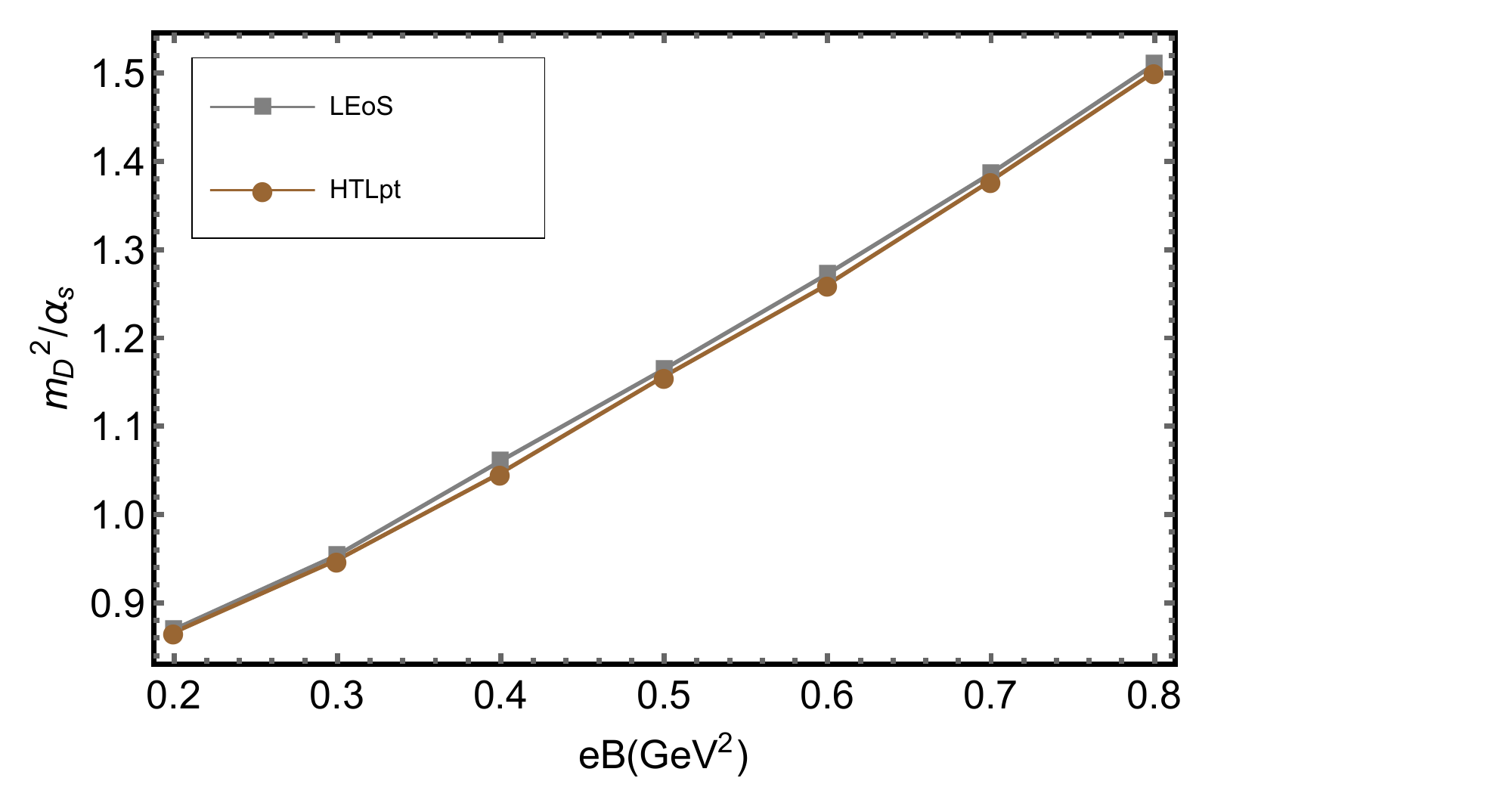}}
 \caption{(Left panel) Temperature behavior of the ratio of Debye mass 
 to coupling constant for different Landau levels 
 at $\mid eB\mid=0.3$ GeV$^2$.(Right panel) 
 Behavior of $m_{D}^{2}/\alpha_{s}$ at T=0.25 GeV$^2$ with different magnetic field. }
  \label{f1}
\end{figure*}
\section{Effective description of thermal relaxation and bulk viscosity in strong magnetic field}
Green-Kubo formula is employed to estimate the bulk 
viscosity of the medium both in the presence and 
the absence of the strong magnetic field 
background in the studies~\cite{Kharzeev:2007wb,
Moore:2008ws,Czajka:2017bod,Hattori:2017qih}. 
In this work, we are adopting the kinetic theory approach for 
the analytical calculation of $\zeta_{eff}$ in the strong magnetic field, 
in which we need to 
start from the relativistic transport equation.  
The strong magnetic field limit,  $T^{2}\ll eB$  has been  considered for computing various quantities under consideration in LLL.
The contributions from higher Landau 
levels are negligible (proportional to $e^{-\frac{\sqrt{eB}}{T}}$) in the regime. Now, for the weaker magnetic fields,  going beyond LLL might  help in understanding the  impact of the magnitude of the field on the 
transport coefficients.  A full computation in the weak field domain will also require 
computation of the quark/antiquark propagators under the same approximation and is beyond the scope of the present work.  The formalism for the estimation of 
effective bulk viscosity includes the 
quasiparticle modeling of the system followed by the 
estimation of the thermal relaxation time of the process.

\subsection{ EQPM in the strong magnetic field}

EQPM describes the hot QCD medium effects with temperature dependent 
effective fugacities - quasigluon and quasiquark/antiquark fugacities, 
$z_{g}$ and $z_{q}$ respectively~\cite{Chandra:2008kz}. 
Various quasiparticle
models encode the medium effects, $viz.$, effective masses 
with Polyakov loop~\cite{D'Elia:97}, NJL and PNJL based 
quasiparticle models~\cite{Dumitru}, self-consistent and
single parameter quasiparticle models~\cite{Bannur:2006js}
and recently proposed quasiparticle models based on the
Gribov-Zwanziger (GZ) quantization, leading to a nontrivial
IR-improved dispersion relation in terms of the Gribov
parameter~\cite{Su:2014rma,zwig,Bandyopadhyay:2015wua}.
EQPM encodes the medium effects as EoS dependence 
of the distribution functions, enters 
through the effective fugacities.

 Here, we consider the recent (2+1) flavor lattice QCD EoS 
(LEoS)~\cite{Cheng:2007jq} and 3-loop HTL perturbative (HTLpt) 
EOS~\cite{Haque, Andersen}. The 3-loop HTLpt EOS has recently 
been computed by N. Haque $et, al$. which is very close to the 
recent lattice results~\cite{Borsanyi, Haque:2014}. These EoSs 
have been carefully embedded in $z_{q}$ and $z_{q}$ for 
both isotropic and to anisotropic hot QCD medium
~\cite{Jamal:2017dqs,Kumar:2017bja}. $z_{q}$ 
and $z_{q}$ have complicated temperature dependence as discussed 
in Ref.~\cite{Agotiya:2016bqr}.   

We have extended the EQPM in the presence 
of magnetic field ($\vec{B}=B\hat{z}$)~\cite{Kurian:2017yxj} 
in which quasi-quark/antiquark 
distribution function is given as,
\begin{equation}\label{1}
\bar{f}^l_{q}=\dfrac{z_{q}\exp{(-\beta \sqrt{p_{z}^{2}+m^{2}
+2l\mid q_feB\mid})}}{1+ z_{q}\exp{( -\beta 
\sqrt{p_{z}^{2}+m^{2}+2l\mid q_feB\mid} )}},
\end{equation}    
where $E^l_{p}=\sqrt{p_{z}^{2}+m^{2}+2l\mid q_feB\mid}$ is the Landau 
energy eigenvalue and $q_{f}e$ is the fractional charge of 
quarks. $l=0,1,2,..$ is the order of the energy levels. 
Since dispersion relation of electrically neutral gluon 
remain intact in strong magnetic field background, the quasigluon
distribution function remains as,
\begin{equation}\label{2}
\bar{f}_{g}=\dfrac{z_{g}\exp{(-\beta\mid\vec{p}\mid)}}
{1+ z_{g}\exp{( -\beta\mid\vec{p}\mid} )}.
\end{equation}
We are working in units where $k_{B}=1$, $c=1$, 
$\hbar=1$ and hence $\beta=\dfrac{1}{T}$.
The parton distribution functions leads to the dispersion relations,
\begin{equation}\label{3}
\omega^l_{q}=\sqrt{p_z^{2}+m^{2}+2l\mid q_feB\mid}+T^{2}\partial_{T} \ln(z_{q}),
\end{equation}
and
\begin{equation}\label{4}
\omega_{g}=\mid\vec{p}\mid+T^{2}\partial_{T} \ln(z_{g}).
\end{equation}
The physical significance of the effective fugacity comes
in the second term of dispersion relations Eqs.~(\ref{3})
and~(\ref{4}) which corresponds to the collective excitation
of quasipartons. Effects of the magnetic field are entering 
into the system through the dispersion relations and  
the Debye screening mass~\cite{Hattori:2016idp}. 
\subsubsection*{Debye mass and effective coupling in the strong magnetic field with HLL corrections}
EQPM is based on charge renormalization in the hot QCD medium
whereas the effective mass model is motivated from the mass
renormalization of QCD~\cite{Mitra:2017sjo}. Realization of 
this charge renormalization could be 
 related to the estimation of Debye mass from semi-classical 
transport theory. There are several investigations on the 
screening masses of the QGP as a function of the magnetic 
field~\cite{Bandyopadhyay:2016fyd,Bonati:2017uvz,Singh:2017nfa}. 
Employing EQPM, we can compute the screening mass as~\cite{Kurian:2017yxj,Mitra:2017sjo},
\begin{equation}\label{5}
m_{D}^{2}=-4\pi\alpha_{s}\int{\dfrac{d^{3}\vec{p}}{(2\pi)^{3}}
\dfrac{d}{d\vec{p}}(2N_{c}\bar{f}_{g}+ N_{f}(\bar{f}^l_{q}+\bar{f}^l_{\bar{q}}))},
\end{equation}
where $\bar{f}^l_{q}$ and $\bar{f}_g$ is the quasiparton distribution 
functions as defined in Eqs.~(\ref{1}) and~(\ref{2}), $\alpha_{s}(T)$
is the running coupling constant at finite temperature taken from
2-loop QCD gauge coupling constants~\cite{Laine:2005ai}. 
Including the effects of HLLs in presence of the strong 
magnetic field $\vec{B}=B\hat{z}$, $m_D$ for quarks and antiquarks becomes,
\begin{equation}\label{6}
m_{D}^{2}=\dfrac{4\alpha_{s}}{T}\dfrac{\mid q_feB\mid}{\pi}
\int_{0}^{\infty}{\sum_{l=0}^{\infty}dp_z(2-\delta_{l0})\bar{f}^l_q(1-\bar{f}^l_q)},
\end{equation}
in which integration phase factor due to dimensional reduction
in the strong field~\cite{Bruckmann:2017pft,Tawfik:2015apa,
Gusynin:1995nb} can be represented as, 
\begin{equation}\label{7}
\int{\dfrac{d^{3}p}{(2\pi)^{3}}}\rightarrow
\dfrac{\mid q_feB\mid}{2\pi}\sum_{l=0}^{\infty}\int{\dfrac{dp_{z}}{2\pi}}(2-\delta_{0l}).
\end{equation}
After performing the momentum integral Eq.~(\ref{5}) 
using Eq.~(\ref{1}) we obtain,
\begin{align}\label{8}
(m_{D}^{2}/\alpha_{s})&= \dfrac{24T^{2}}{\pi}PolyLog[2,z_{g}]
+\dfrac{12\mid q_feB\mid}{\pi}\left(  \dfrac{z_{q}}{1+z_{q}}\right)\nonumber\\
&+ \dfrac{8}{T}\dfrac{\mid q_feB\mid}{\pi}
\int_{0}^{\infty}{\sum_{l=1}^{\infty}dp_z \bar{f}^l_q(1-\bar{f}^l_q)}.
\end{align}
We have plotted the ratio of Debye mass to running coupling 
constant ratio at $\mid eB\mid=0.3$ GeV$^2$ as a function of
temperature for different Landau levels in Fig.~\ref{f1}. 
For the chosen temperature range we are focusing up to $l=3$ Landau level. The
contribution from HLLs beyond $l=3$ is negligible for the given temperature range.  
Since the occupation
in HLLs is exponentially suppressed by 
$\exp{-(\dfrac{\mid eB\mid}{T})}$, 
the effect of HLLs will be significant for higher temperature 
ranges. For ideal EoS $z_{q,g}=1$ 
(ultra-relativistic non-interacting quarks and gluons), 
the definition of Debye mass can rewrite as,
\begin{align}\label{8.1}
(m^{2}_{D})_{Ideal}&=4\pi\alpha_{s}(T)[T^{2}+\dfrac{3\mid q_feB\mid}{2\pi^{2}}\nonumber\\
&+\dfrac{2}{T}\dfrac{\mid q_feB\mid}{\pi^2}
\int_{0}^{\infty}{\sum_{l=1}^{\infty}dp_z \bar{n}^l_q(1-\bar{n}^l_q)}],
\end{align}
with $\bar{n}^l_{q}=\dfrac{1}{\exp{(\beta E^l_p)}+ 1}$
From Eqs.~(\ref{8}) and~(\ref{8.1}), including HLLs we can 
define the effective running coupling constant $\alpha^l_{eff}(T,z_q,z_g,\mid eB\mid)$ so that, 
\begin{align}
m^{2}_{D}=\dfrac{\alpha^l_{eff}}{\alpha_s}{m^{2}_{D}}_{Ideal}.
\end{align}
Therefore, 
\begin{align}
m^{2}_{D}&= 4\pi \alpha^l_{eff}(T,z_{q},z_{g})[T^{2}+
\dfrac{3\mid q_feB\mid}{2\pi^{2}}\nonumber\\
&+\dfrac{2}{T}\dfrac{\mid q_feB\mid}{\pi^2}
\int_{0}^{\infty}{\sum_{l=1}^{\infty}
dp_z \bar{n}^l_q(1-\bar{n}^l_q)}],
\end{align}
Now , $ \alpha^l_{eff}$ 
can be expressed as,
\begin{figure}[h]
  \subfloat{\includegraphics[height=7.5cm,width=8.2cm]{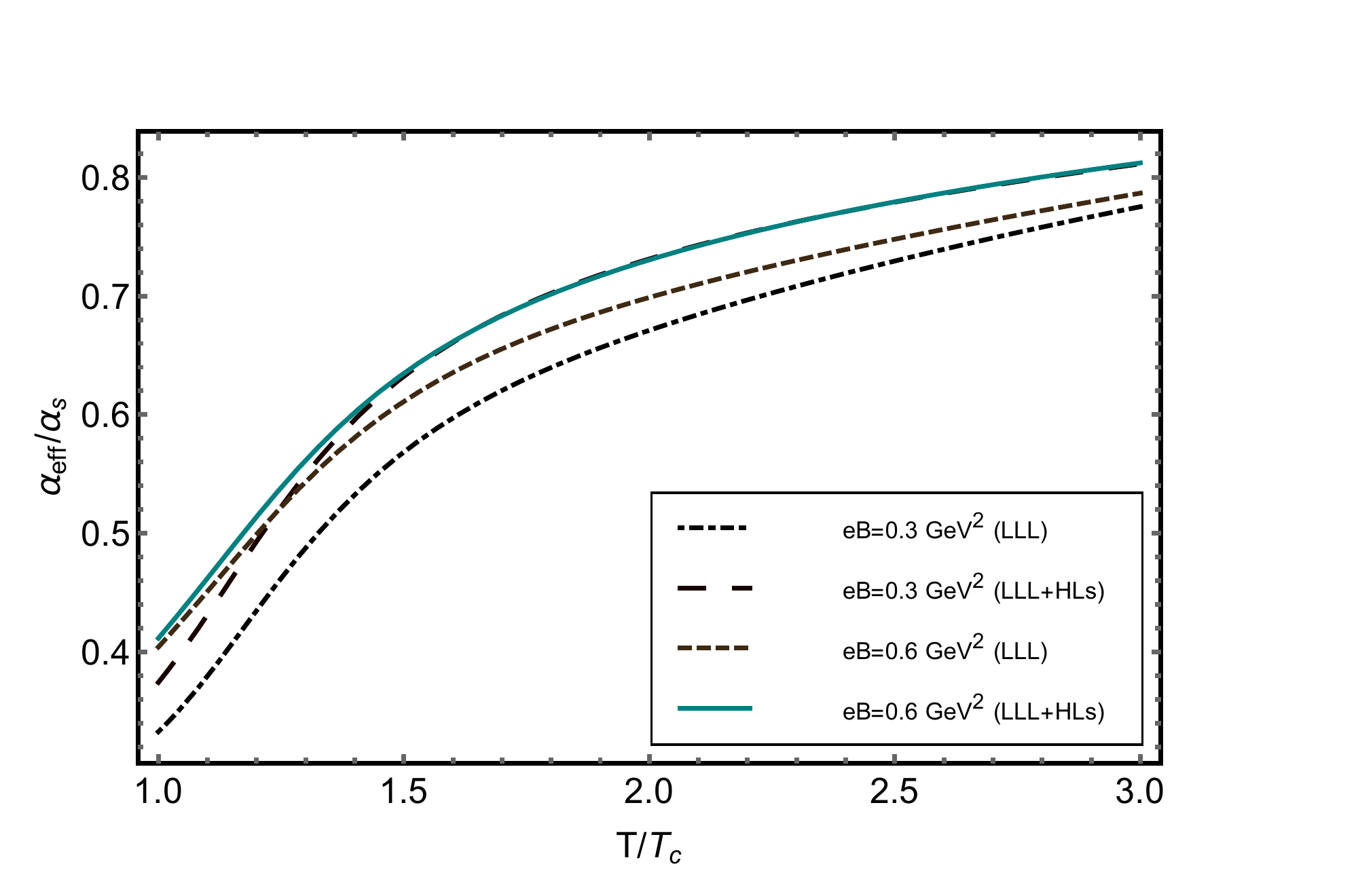}}
\caption{(color online) The effective coupling constant in the strong 
magnetic field with HLLs corrections. }
\label{f2}
\end{figure}
\begin{align}\label{9}
\dfrac{\alpha^l_{eff}}{\alpha_{s}}=&
\dfrac{ \dfrac{6T^{2}}{\pi^{2}}
PolyLog[2,z_{g}]+\dfrac{3\mid q_feB\mid}
{\pi^{2}}\dfrac{z_{q}}{(1+z_{q})}}
{\left( T^{2}+\dfrac{3\mid q_feB\mid}{2\pi^{2}}+h(T,\mid eB\mid)\right)}\nonumber\\
&+ \dfrac{\dfrac{2}{T}\dfrac{\mid q_feB\mid}{\pi^2}
\int_{0}^{\infty}{\sum_{l=1}^{\infty}dp_z 
\bar{f}^l_q(1-\bar{f}^l_q)}}{\left( T^{2}
+\dfrac{3\mid q_feB\mid}{2\pi^{2}}+h(T,\mid eB\mid)\right)},
\end{align}
where $h(T,\mid eB\mid)=\dfrac{2}{T}
\dfrac{\mid q_feB\mid}{\pi^2}\int_{0}^{\infty}
{\sum_{l=1}^{\infty}dp_z \bar{n}^l_q(1-\bar{n}^l_q)}$.

 For LLL quarks Eq.~(\ref{9}) reduced to
\begin{equation}\label{9.1}
  {\dfrac{ \alpha^0_{eff} }{\alpha_{s}}}  =\dfrac{\left( \dfrac
{6T^{2}}{\pi^{2}}PolyLog[2,z_{g}]+\dfrac{3\mid q_feB\mid}
{\pi^{2}}\dfrac{z_{q}}{(1+z_{q})}\right)}{\left
( T^{2}+\dfrac{3\mid q_feB\mid}{2\pi^{2}}\right)}.
\end{equation}
The temperature behavior of $\frac{\alpha^l_{eff}}{\alpha_{s}}$ with 
HLLs corrections are depicted in the Fig.~\ref{f2}. 
As expected, asymptotically the ratio approaches unity. Dominant contribution 
of $\alpha^l_{eff}$ comes from the LLL, where the HLLs gives the 
higher order corrections. More interestingly, including HLLs 
$\frac{\alpha^l_{eff}}{\alpha_{s}}$ are almost identical for 
$\mid eB\mid=0.3$ GeV$^2$ and  $\mid eB\mid=0.6$ GeV$^2$, 
which implies the weaker dependence of the strength of 
magnetic field on $\frac{\alpha^l_{eff}}{\alpha_{s}}$. 
The ratio is showing a small but quantitative 
change with the HLLs corrections. Hence, these corrections 
are significant in the estimation of the
bulk viscosity in the 
strong field.

\subsection{Thermal relaxation in strong magnetic field}
The microscopic interactions, which are the dynamical inputs 
of the bulk viscosity, are incorporated through the thermal 
relaxation time ($\tau_{eff}$). The focus of this work is on 
the dominant 1 $\rightarrow$ 2 processes (gluon to quark-antiquark pair).  
The relaxation time,
$\tau_{eff}$ can be defined from the relativistic transport 
equation of quasiparton distribution functions for the process 
$ k\rightarrow p+p^{'} $ in strong magnetic field $\vec{B}=B\hat{z}$ as,
\begin{equation}\label{10}
\dfrac{d f^l_{q}}{d t}=C(f^l_{q})=-\dfrac{\delta f^l_q}{\tau_{eff}}.
\end{equation}
The quantity $\delta f^l_{q}$ is the non-equilibrium part of
the distribution function of quasiquark/antiquark, 
\begin{equation}\label{10.1}
f^l_{q}(p_{z})= \bar{f}_{q}^{l}+\delta f^l_{q},
\end{equation}
and given by
\begin{equation}\label{11}
\delta f^l_q= \beta \bar{f}^l_{q}({p_{z}})(1-\bar{f}^l_{q}({p_{z}}))
 \chi_{q}(p_{z}),
\end{equation}
where $\chi(p_{z})$ is the response function (primed notation 
for antiquark). Here, $C(f^l_{q})$ is the collision integral which 
quantifies the rate of change of distribution function. 
In strong magnetic field background, 
the collision integral for 1 $\rightarrow$ 2 processes have the 
following form~\cite{Kurian:2017yxj}  
\begin{align}\label{12}
C(f^l_{q})&=\alpha^l_{eff}C_{2}m^2\int_{-\infty}^{\infty}
{\dfrac{dp^{'}_{z}}{\omega^l_{p}\omega^l_{p^{'}}}}
\beta \bar{f}_{q}^{l}(E^l_{p^{'}})\bar{f}_{q}^{l}(E^l_{p})\nonumber\\
&\times ( 1+\bar{f}_{g}(E^l_{p}+E^l_{p^{'}})
\left( \chi_{q}(p_{z}^{'})-\chi_{q}(p_{z})\right), 
\end{align}
where $C_2$ is the Casimir factor and $\alpha^l_{eff}$ is the 
effective coupling constant which encoded the EoS dependence. 
$\omega^l_p$ is the single quark energy as defined in Eq.~(\ref{3}).
The response $\chi$ for quark and antiquark in the strong magnetic 
 field has opposite sign (since their charges are opposite). 
 This implies that $\chi_{q}(p_{z}^{'})$ is an odd function 
 as described in~\cite{Hattori:2016lqx} within LLL approximation. 
 Since the Landau 
 levels enters as $E^l=\sqrt{p_z^2+m^2+2l\mid eB\mid}$ 
 in the dispersion relations and distribution functions, 
 the odd nature of $\chi(p_{z}^{'})$ is completely independent on the order of LL. Hence we have,

\begin{align}\label{12.1}
C(f^l_{q})&=-\chi_{q}(p_{z})\alpha^l_{eff}C_{2}m^2\beta\nonumber\\
&\times\int_{-\infty}^{\infty}
{\dfrac{dp^{'}_{z}}{\omega^l_{p}\omega^l_{p^{'}}}}
\bar{f}_{q}^{l}(E^l_{p^{'}})\bar{f}_{q}^{l}(E^l_{p})
 ( 1+\bar{f}_{g}(E^l_{p}+E^l_{p^{'}}). 
\end{align}
Thermal relaxation time $\tau_{eff}$, which is the inverse 
of the quark damping rate $\Gamma_{eff}$, can be obtained from 
Eqs.~(\ref{10})~(\ref{11}) and~(\ref{12.1}) as,
\begin{align}\label{13}
\tau_{eff}^{-1}&\equiv \Gamma_{eff}\nonumber\\
&=\dfrac{\alpha^l_{eff}C_{2} m^2}{\omega_p (1-\bar{f}_{q}^{l})}
\int{\dfrac{dp^{'}_{z}}{\omega^l_{p^{'}}}} 
\bar{f}_{q}^{l}(E^l_{p^{'}})( 1+\bar{f}_{g}(E^l_{p}+E^l_{p^{'}})). 
\end{align}
\begin{figure}[h]
  \subfloat{\includegraphics[height=6.5cm,width=8.5cm]{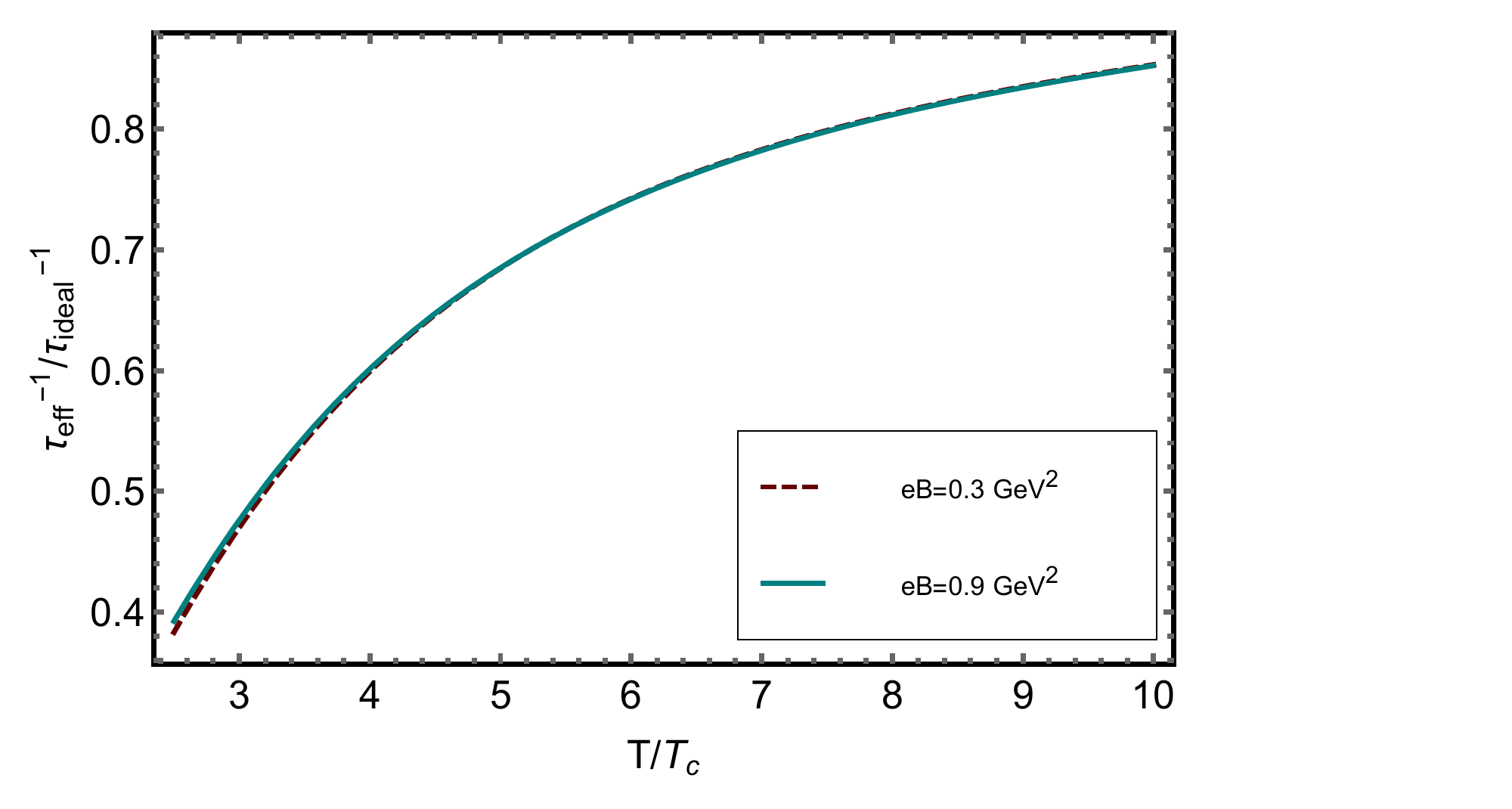}}
\caption{(color online) Dependence of EoS on the thermal relaxation time 
in the strong magnetic field for the LLL quarks with $<p_z>=T$.}
\label{f3}
\end{figure}
Being motivated by the recent work Ref.~\cite{Hattori:2016cnt,Hattori:2016lqx}, 
we constrained our calculation in the regime in which the dominant 
contribution comes from the quarks of the momentum of order $T$.
Hence, the energy of quarks $E_q\sim T$ and 
this makes the gluon energy $E_q+E_{\bar{q}}\sim T$, 
where $E_{\bar{q}}$ is the quark energy. Hence, we have $p_{z}^{'}\ll T$ 
or $\frac{p_{z}^{'}}{T}\sim 0$~\cite{Hattori:2016cnt,Hattori:2016lqx}. 
Solving the integral in Eq.(\ref{13}) 
within this assumptions gives the logarithmic factor. Finally, 
we obtain the momentum dependent thermal relaxation time 
$\tau_{eff}(p_z,z_{q/g},\mid eB\mid)$ from the extended EQPM as,
\begin{equation}\label{14}
\tau_{eff}^{-1}=\dfrac{2\alpha^l_{eff}C_{2} m^2}{\omega^l_p 
(1-\bar{f}_{q}^{l})}\dfrac{z_q}{(z_{q}+1)}
( 1+\bar{f}_{g}(E^l_{p}))\ln{(T/m)}. 
\end{equation}
Impact of the hot QCD medium effects on the relaxation time can be explored by comparing it with the case where the  hot QCD/QGP is described as the free ultra-relativistic gas of quarks and gluons, as done in~\cite{Hattori:2016lqx}. This could  be described  by choosing  $z_{g/q} = 1$, and in that case, the  relaxation time reduces to
\begin{equation}\label{15}
\tau_{ideal}^{-1}=\dfrac{\alpha_{s}C_{2} m^2}{E^l_p 
(1-\bar{n}_{q}^{l})}
( 1+\bar{n}_{g}(E^l_{p}))\ln{(T/m)}, 
\end{equation}
with $\bar{n}_{q}^{l}=\dfrac{1}{(e^{\beta E^l_p}+ 1)}$ and 
$\bar{n}_{g}=\dfrac{1}{(e^{\beta E^l_p}- 1)}$ for ideal 
fermions and bosons respectively.

Since the dominant charge carriers have momenta in the order of T, we are employing
$<p_z>=T$ for the comparison of $\tau_{eff}$ with $\tau_{ideal}$ 
to investigate the EoS dependence. Note that the momentum dependence 
of the relaxation time is significant in the estimation of 
bulk viscosity. Therefore, while computing the bulk viscosity, 
the momentum dependent thermal relaxation time as defined in Eq.(\ref{14}) is employed.
Here, we plotted the 
temperature variation of $\frac{\tau_{eff}^{-1}}{\tau_{ideal}^{-1}}$ with $<p_z>=T$
for the ground state quarks ($l=0$)
at $\mid eB\mid=0.3$ GeV$^2$ and 
$\mid eB\mid=0.9$ GeV$^2$ in Fig~\ref{f3}. Hot medium 
effects are identical for the system under consideration 
irrespective of the magnitude of the magnetic field. EoS effects in 
relaxation time are embedded in Eq.~(\ref{13}) through the 
quasiparton distribution function and the effective coupling 
defined in Eq.~(\ref{9}). Since $\alpha^l_{eff}$ is lower than 
$\alpha_s$ at the lower temperature, $\tau_{eff}^{-1}$ to  
$\tau_{ideal}^{-1}$ ratio gives lower value in that temperature 
range.

HLLs corrections are entering through Landau dispersion relation
in the quark distribution function. Effect of higher levels 
in the effective coupling is understood from Eq.~(\ref{9}).
 The effective thermal relaxation time controls the behavior 
of bulk viscosity critically.
\subsection{Bulk viscosity from the relaxation time approximation}
We investigated the bulk viscosity of 
perturbative QCD in 
the strong magnetic field $\vec{B}=B\hat{z}$ by adopting the EQPM. 
for the dominant 1 $\rightarrow$ 2 
processes. Dynamics of the system is described 
by the Boltzmann equation for the quasiquark distribution function,
\begin{equation}\label{16}
\left( \partial_t+v_z\partial_z\right)   f_q^l(p_z,t,z)=C(f^l_{q})
=-\dfrac{\delta f^l_q}{\tau_{eff}},
\end{equation}
where $C(f^l_{q})$ is the collision integral Eq.~(\ref{12}) and the 
longitudinal velocity $v_z\equiv \frac{\partial \omega_p^l}{\partial p_z}=\frac{p_z}{E_p}$. 
Equilibrium distribution function is defined as,
\begin{equation}\label{17}
\bar{f}^l_{q}=\dfrac{1}{\left( z_{q}^{-1}\exp{(-\beta( E^l_p-p_zv_z))}+1\right) },
\end{equation}
in the presence of the flow $u_z$. For $u_z=0$, $\bar{f}^l_{q}$ reduces 
to Eq.~(\ref{1}). We consider the linear response regime of the 
Boltzmann equation in which $u_z$ and $\delta f^l_{q}$ are assumed to be small, 
with appropriate collision integral to solve  $\delta f^l_{q}$.
The system in equilibrium is disturbed by an expansion 
in the direction of magnetic field, which gives the change in 
pressure ($\delta P_L$). 
Bulk viscosity is defined as~\cite{Hattori:2017qih},
\begin{equation}\label{18}
\delta P_L=-3\zeta_{eff} \Theta,
\end{equation}
with $\Theta(z)\equiv \partial_z u_z$, which defines the magnitude 
of expansion. We investigated the QCD thermodynamic quantities such 
as pressure, energy density, entropy density and the speed of sound 
in the strong magnetic field
using the extended EQPM~\cite{Kurian:2017yxj}. With LLL approximation, 
longitudinal pressure (in the direction of $\vec{B}$) is obtained 
from the fundamental thermodynamic definition, 
\begin{align}\label{ad1}
 P_L=&\sum_f\dfrac{\mid eq_{f}B\mid }{2\pi}\dfrac{1}{2\pi}2N_c\nonumber\\
 &\times\int_{-\infty}^{\infty}{dp_z\ln{(1+z_q
 \exp{(-\beta\sqrt{p_z^2+m^2})})}}.
\end{align}
Longitudinal pressure end up as, 
\begin{equation}\label{19}
 P_L=\sum_f\dfrac{\mid eq_{f}B\mid }{\pi^2}N_c
 \int_{0}^{\infty}{dp_z\dfrac{p_z^2}{E^0_p}\bar{f}^0_q},
\end{equation}
where $\bar{f}^0_q$ is the momentum distribution of lowest Landau quarks ($l=0$). 
Similarly, energy density of the quarks is defined as
\begin{equation}\label{20}
 \varepsilon_L=\sum_f\dfrac{\mid eq_{f}B\mid }{\pi^2}N_c
 \int_{0}^{\infty}{dp_z\dfrac{{(\omega^0_p)}^2}{\omega^0_p}\bar{f}^0_q},
\end{equation}
in which $\omega^0_p$ is the single particle energy for LLL quarks.
The integral can be expressed in terms of $PolyLog$ functions. 
Change in longitudinal pressure leads to the bulk viscosity in 
the direction of magnetic field as given in Eq.~(\ref{18}) and hence,
\begin{equation}\label{21}
\zeta_{eff} =\sum_f-\dfrac{1}{3\Theta}\dfrac{\mid eq_{f}B\mid }
{\pi^2}N_c\int_{0}^{\infty}{dp_z\dfrac{p_z^2}{E^0_p}\delta f^0_q}.
\end{equation}
However, even when $\delta f^l_{q}=0$ there will be a change in 
pressure since the temperature $(\beta\equiv\beta (t))$ decreases 
in time due to the expansion. This can be directly related to the 
Landau-Lifshitz condition for the stress-energy tensor in the 
calculation of the bulk viscosity without magnetic field~\cite{Chakraborty:2010fr}. 
We subtract this effect as in Ref.~\cite{Hattori:2017qih,Arnold:2006fz}, 
and we have
\begin{equation}\label{22}
\delta P\rightarrow\delta \bar{P}_L\equiv \delta (P_L-\Omega \varepsilon_L),
\end{equation}
with $\Omega\equiv\dfrac{\partial P_L}{\partial\varepsilon_L}
=\dfrac{\partial P_L/\partial T}{\partial\varepsilon_L/\partial T}$. 
To solve this, we have used EQPM definition of pressure and 
energy density  
in strong magnetic field as in Eqs.~(\ref{19}) and~(\ref{20}), 
\begin{align}\label{23}
\Omega&= \lbrace -\mid eB\mid\dfrac{2T}
{\pi^{2}}\bar{\nu}_qPolyLog[2,-z_{q}]\nonumber\\
&+\mid eB\mid\dfrac{T^{2}}{\pi^{2}}\bar{\nu}_q
\ln(1+z_{q})\left(\partial_{T}\ln z_{q}\right)\rbrace /\nonumber\\
&\lbrace -\dfrac{4\mid eB\mid T}
{\pi^{2}}\bar{\nu}_qPolyLog[2,-z_{q}]\nonumber\\
&+5\mid eB\mid\left( T^{2}\partial_{T}\ln z_{q}\right)
\dfrac{1}{\pi^{2}}\bar{\nu}_q\ln(1+z_{q})\nonumber\\
&+\mid eB\mid T^{2}\left(\partial_{T}
\ln z_{q}\right)^{2}\dfrac{T}{\pi^{2}}\bar{\nu}_q
\dfrac{z_{q}}{1+z_{q}}\nonumber\\
&+\mid eB\mid T^{2}\left( \partial^{2}_{T}
\ln z_{q}\right)\dfrac{T}{\pi^{2}}\bar{\nu}_q\ln(1+z_{q})\rbrace ,
\end{align}

where $\bar{\nu}_{q}=\sum_{f} 2N_{c}\mid q_{f}\mid$ in the presence 
of magnetic field. Also, we need to evaluate the change in equilibrium 
distribution function $\delta f^l_{q}$ for the calculation of 
$\delta \bar{P}_L$. Considering the linear response regime of the 
Boltzmann equation Eq.~(\ref{16}) with the distribution 
function as Eq.~(\ref{17}), we have
\begin{align}\label{24}
\left( \partial_t+v_z\partial_z\right)  f^0_q(p_z,t,z)
=&-[(E^0_p+T^2\partial_T\ln z_q)\partial_t\beta\nonumber\\
&-\beta v_zp_z\Theta(z)]\bar{f}^0_{q}(\bar{f}^0_{q}-1).
\end{align}
\begin{figure}[h]
  \subfloat{\includegraphics[height=6.1cm,width=9.3cm]{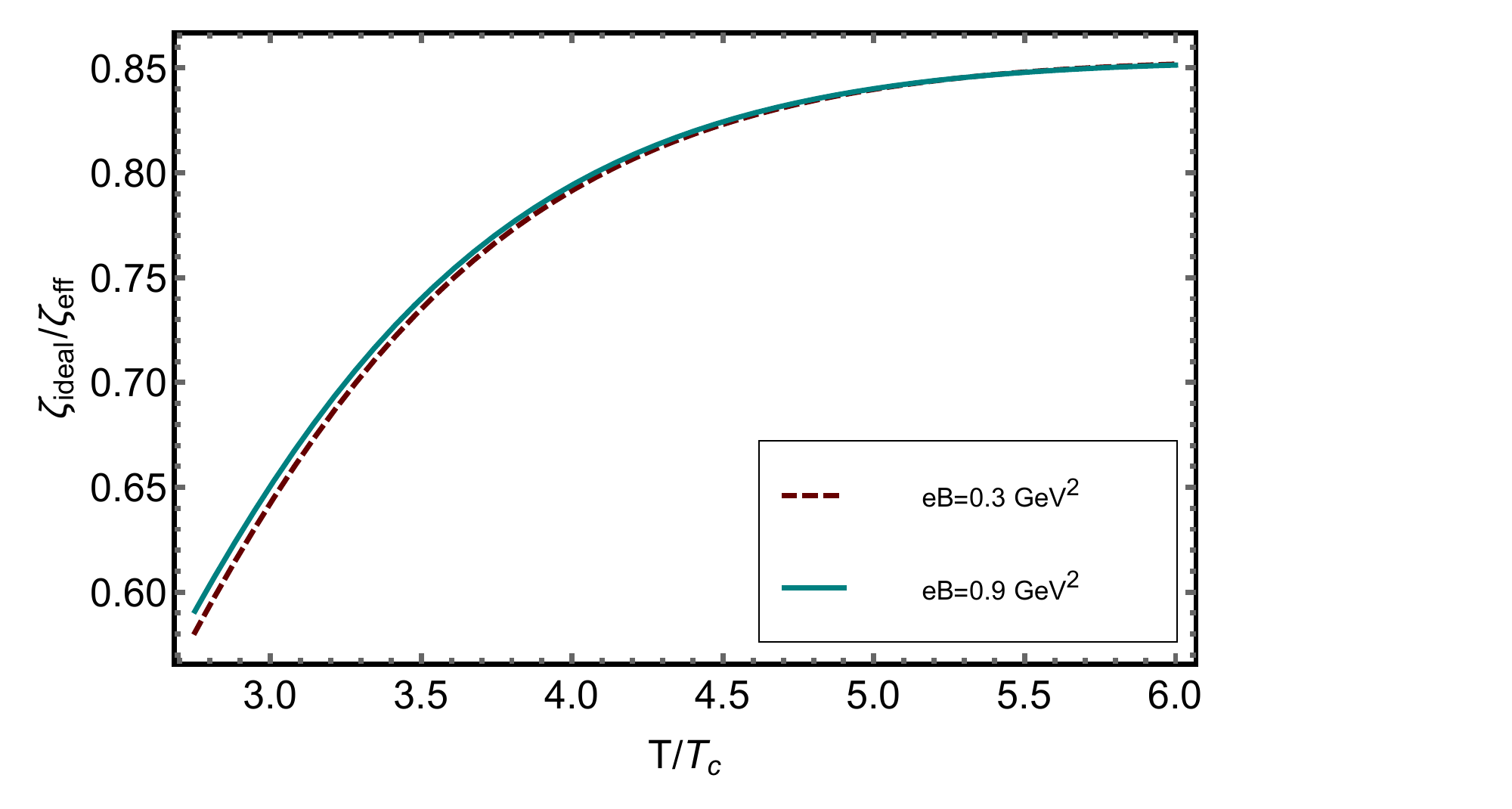}}
\caption{(color online) Dependence of EoS on the bulk viscosity in 
the strong magnetic field with LLL approximation.}
\label{f5}
\end{figure}
Here, $z_q(\frac{T}{T_c})$ and $\beta(T)$ are 
functions of time since temperature 
changes with expansion. Detailed calculations are 
shown in the Appendix A. 
In the relaxation time approximation, we can directly connect the 
relaxation time $\tau_{eff}$ with the collision integral 
$C(f^l_{q})$ as shown in Eq.~(\ref{16}). Therefore, Eq.~(\ref{24}) 
becomes, 
\begin{figure*}
 \centering
 \subfloat{\includegraphics[height=6.8cm,width=8.5cm]{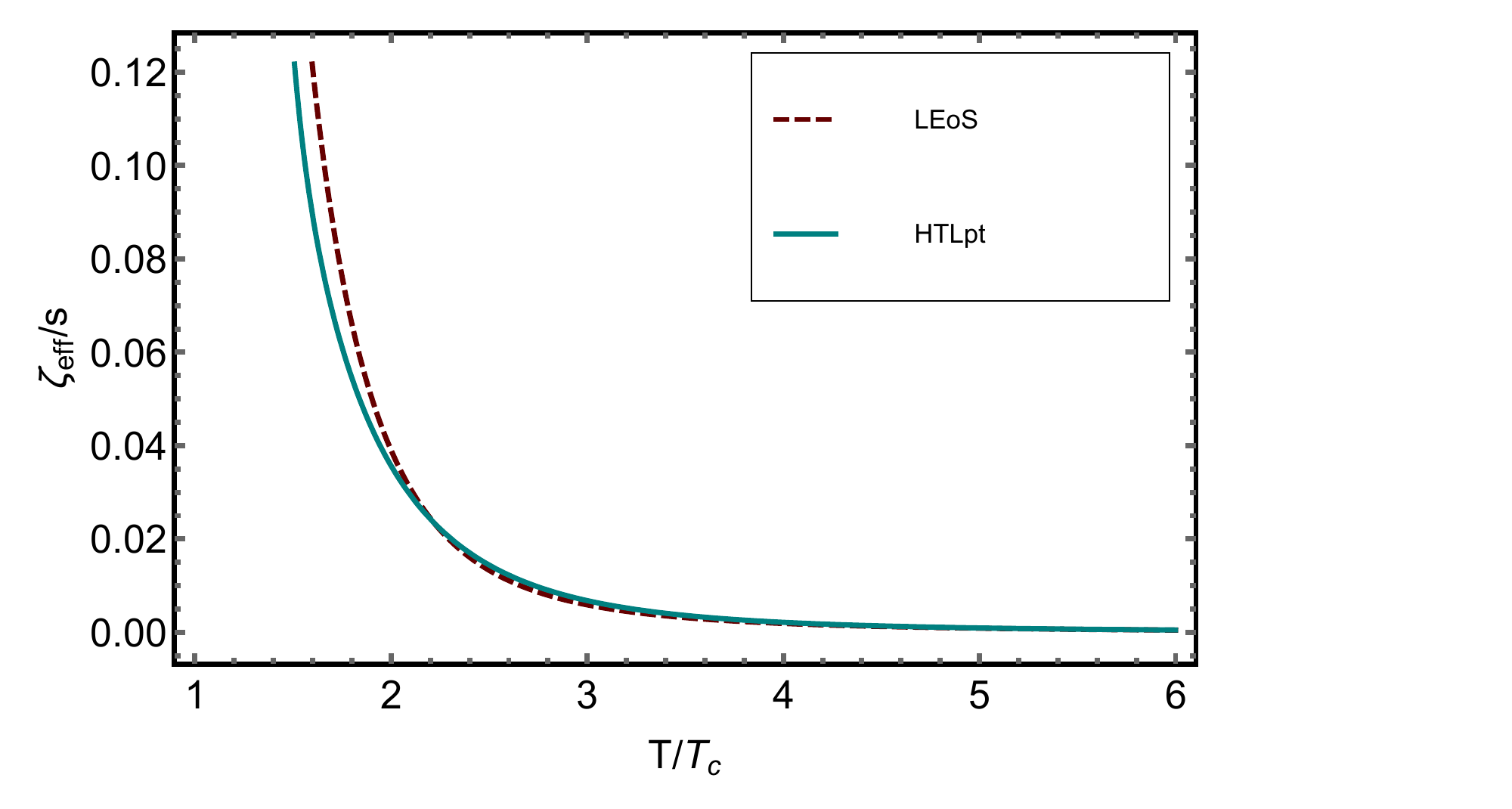}}
% \hspace{1 cm}
 \subfloat{\includegraphics[height=6.80cm,width=8.50cm]{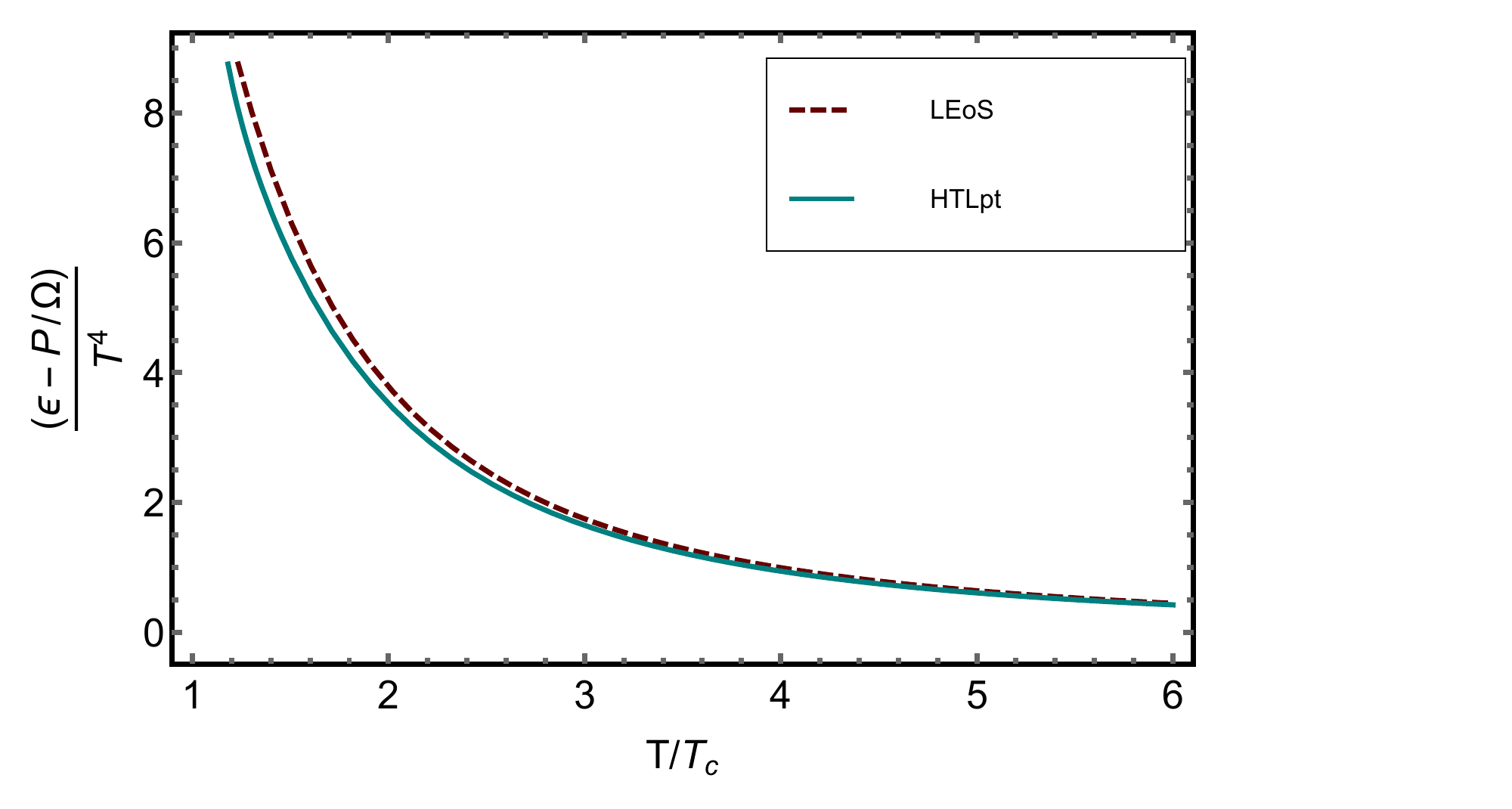}}
 \caption{ Temperature behavior of ratio of bulk viscosity to entropy 
 (left panel) and $\dfrac{(\varepsilon-P/\Omega)}{T^4}$ (right panel) 
 at $\mid eB\mid=0.3$  GeV$^2$ with LLL approximation.}
 \label{f4}
\end{figure*}
\begin{equation}\label{25}
\delta f^0_{q}=-\tau_{eff}\beta \bar{f}^0_{q}(\bar{f}^0_{q}-1)
\Theta(z) (\omega^0_p\Omega-v_zp_z),
\end{equation}
where $\partial_t\beta\equiv\beta\Omega\Theta$ as given 
in~\cite{Hattori:2017qih,Arnold:2006fz} 
and $\tau_{eff}$ is the thermal relaxation time {(at $l=0$ in the LLL approximation)
for $1\rightarrow 2$ processes defined in Eq.~(\ref{14}). 
Now we can estimate $\zeta_{eff}$ by direct substitution of 
Eqs.~(\ref{14})~(\ref{25}) and~(\ref{23}) to~(\ref{22}) 
and end up with,
\begin{align}\label{26}
\zeta_{eff} =& \dfrac{1}{3}\dfrac{\mid q_feB\mid}{\pi^2}\dfrac{\beta}{m^2}
\dfrac{(z_q+1)}{z_q}\dfrac{1}{\alpha^0_{eff} C_2\ln (T/m)}\nonumber\\
&\times\int_{0}^{\infty}{\dfrac{dp_z(p_z^2-\Omega
\omega^0_pE^0_p)^2\bar{f}^0_{q}(1-\bar{f}^0_{q})^2}{(\bar{f}_{g}+1)E^0_p}}.
\end{align}
Here, $\bar{f}^0_{q}$ is the quark distribution 
function with $l=0$ level.
Bulk viscosity $\zeta_{eff}$ depends on the behavior the term 
$(p_z^2-\Omega\omega^0_pE^0_p)^2$ along with the momentum distribution 
function and the effective coupling constant. 
\subsection{Bulk viscosity beyond LLL approximation}
 Effect of HLLs on the effective coupling $\alpha^l_{eff}$ and 
thermal relaxation $\tau_{eff}$ are 
defined in the Eq.(\ref{9}) and  Eq.(\ref{13}) respectively. 
Higher order Landau level corrections to the QCD 
thermodynamics (pressure, entropy density etc.) are 
described in our previous work~\cite{Kurian:2017yxj} and utilized in the present work wherever required.
The longitudinal pressure and energy density with HLL corrections have the form,
\begin{equation}\label{26.1}
 P_L=\sum_f\dfrac{\mid eq_{f}B\mid }{\pi^2}N_c
 \int_{0}^{\infty}{dp_z(2-\delta_{0l})\dfrac{p_z^2}{E^l_p}\bar{f}^l_q},
\end{equation} 
and
\begin{equation}\label{26.2}
 \varepsilon_L=\sum_f\dfrac{\mid eq_{f}B\mid }{\pi^2}N_c
 \int_{0}^{\infty}{dp_z(2-\delta_{0l})\dfrac{{(\omega^l_p)}^2}{\omega^l_p}\bar{f}^l_q},
\end{equation}
in which $E^l_{p}=\sqrt{p_{z}^{2}+m^{2}+2l\mid q_feB\mid}$ 
is the Landau levels of order $l$.
The integration phase factor and quasiquark distribution 
function are defined in Eqs.~(\ref{7}) and~(\ref{1}) respectively.
Incorporating these, we can calculate $\bar{\Omega}\equiv\dfrac{\partial P}
{\partial\varepsilon}$ with higher order corrections. 
Finally, the bulk viscosity with higher Landau corrections has the following form,
\begin{align}\label{27}
\zeta_{eff} =&\dfrac{\mid q_feB\mid}{3\pi^2}\sum_{l=0}^{\infty}(2-\delta_{0l})\dfrac{\beta}{m^2}
\dfrac{1}{\alpha^l_{eff} C_2\ln (T/m)}\nonumber\\
&\times\dfrac{(z_q+1)}{z_q}\int_{0}^{\infty}{\dfrac{dp_z(p_z^2-\bar{\Omega}
\omega^l_pE^l_p)^2\bar{f}^l_{q}(1-\bar{f}^l_{q})^2}{(\bar{f}_{g}+1)E^l_p}}.
\end{align}

In transport theory, the viscosity to entropy ratio 
$\zeta_{eff}/s$ has significant importance. The temperature 
behavior and the effects of HLLs on $\zeta_{eff}/s$ are discussed
in the next section. 
  
\section{Results and discussions}

We initiate our discussions with the hot QCD 
medium dependence on the thermal relaxation time $\tau_{eff}$ 
and the effective coupling $\alpha^l_{eff}$. The 
medium dependence on $\alpha^l_{eff}$ and $\tau_{eff}$ 
are explicitly shown in Fig.~\ref{f2} and Fig.~\ref{f3} 
respectively. Thermal relaxation time defined in Eq.~(\ref{14}) 
encoded the microscopic interactions of the system, 
which are the dynamical inputs for the estimation of bulk viscosity.
The hot QCD medium effects embedded through EoS dependence 
on the bulk viscosity of $1\rightarrow 2$ processes can be 
inferred from the Eq.~(\ref{26}). The EoS dependence is 
entering through the quasiparton momentum distribution 
functions along with the effective coupling. 
We plotted the variation of $\zeta_{eff}/\zeta_{ideal}$ 
with $T/T_c$ for $\mid eB\mid=0.3$ GeV$^2$ and $\mid eB\mid=0.9$ GeV$^2$
in Fig.~\ref{f5}. We can see that medium effects are 
weakly depending on the magnitude of magnetic field. $\zeta_{ideal}$, 
bulk viscosity without the medium effects are 
shown in the Ref.~\cite{Hattori:2017qih}. Asymptotically, 
the ratio approaches unity. Hence, the estimation of bulk 
viscosity with quasiparticle modeling agrees with 
the order of magnitude of the results in Ref.~\cite{Hattori:2017qih}
at high temperature.
\begin{figure}[h]
  \subfloat{\includegraphics[height=6.81cm,width=8.2cm]{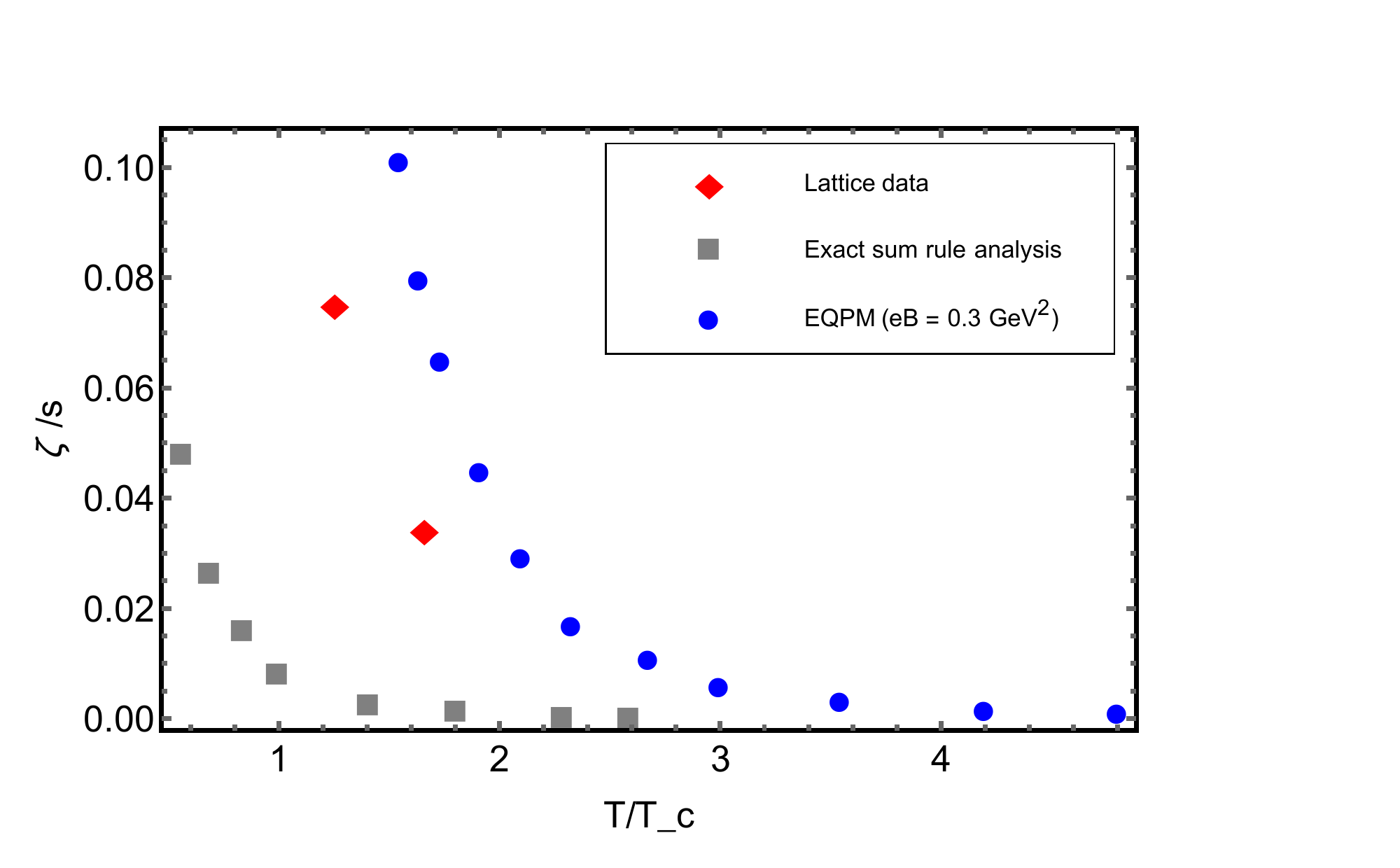}}
\caption{Comparison of the temperature behavior of $\zeta/s$ for 
the $1\rightarrow 2$ processes at $\mid eB\mid=0.3$ GeV$^2$ with Lattice data~\cite{Meyer:2007dy,prasanth:2018} and sum rule analysis~\cite{Karsch:2007jc} in the absence of magnetic field.}
\label{f5.1}
\end{figure}

 Next, we present the temperature behavior of bulk viscosity 
to entropy ratio for $1\rightarrow 2$ 
process in strong magnetic field. Explicit dependence 
of temperature on $\zeta_{eff}/s$ is shown in Eqs.~(\ref{26}) and~(\ref{27}). 
The Eq.~(\ref{14}) shows that
the coupling constant $\alpha$ entering through the 
relaxation time (and hence bulk viscosity) of $1\rightarrow 2$ processes  as $1/{\alpha}$ 
whereas for $2\rightarrow 2$ processes as $1/{\alpha^2}$.
In Fig.~\ref{f4}, we have depicted $\zeta_{eff}/s$  
in the presence of magnetic field 
as a function of $T/T_c$ for both the EoS in LLL approximation. 
The behavior of bulk viscosity depends on 
the $\Omega$. The temperature behavior of 
$(\epsilon-\frac{P}{\Omega})/T^4$ is shown in Fig.~\ref{f4}. 
This term is significantly important in the  Eq.~(\ref{26}) of $\zeta_{eff}/s$. 
The higher value of $\zeta_{eff}/s$ near to the transition temperature $T_c$ 
is due the term  $(\epsilon-\frac{P}{\Omega})/T^4$.  
At very high temperature $\zeta_{eff}/s$ approaches to zero.

\begin{figure}[h]
\vspace{-10mm}
  \subfloat{\includegraphics[height=7.1cm,width=8.cm]{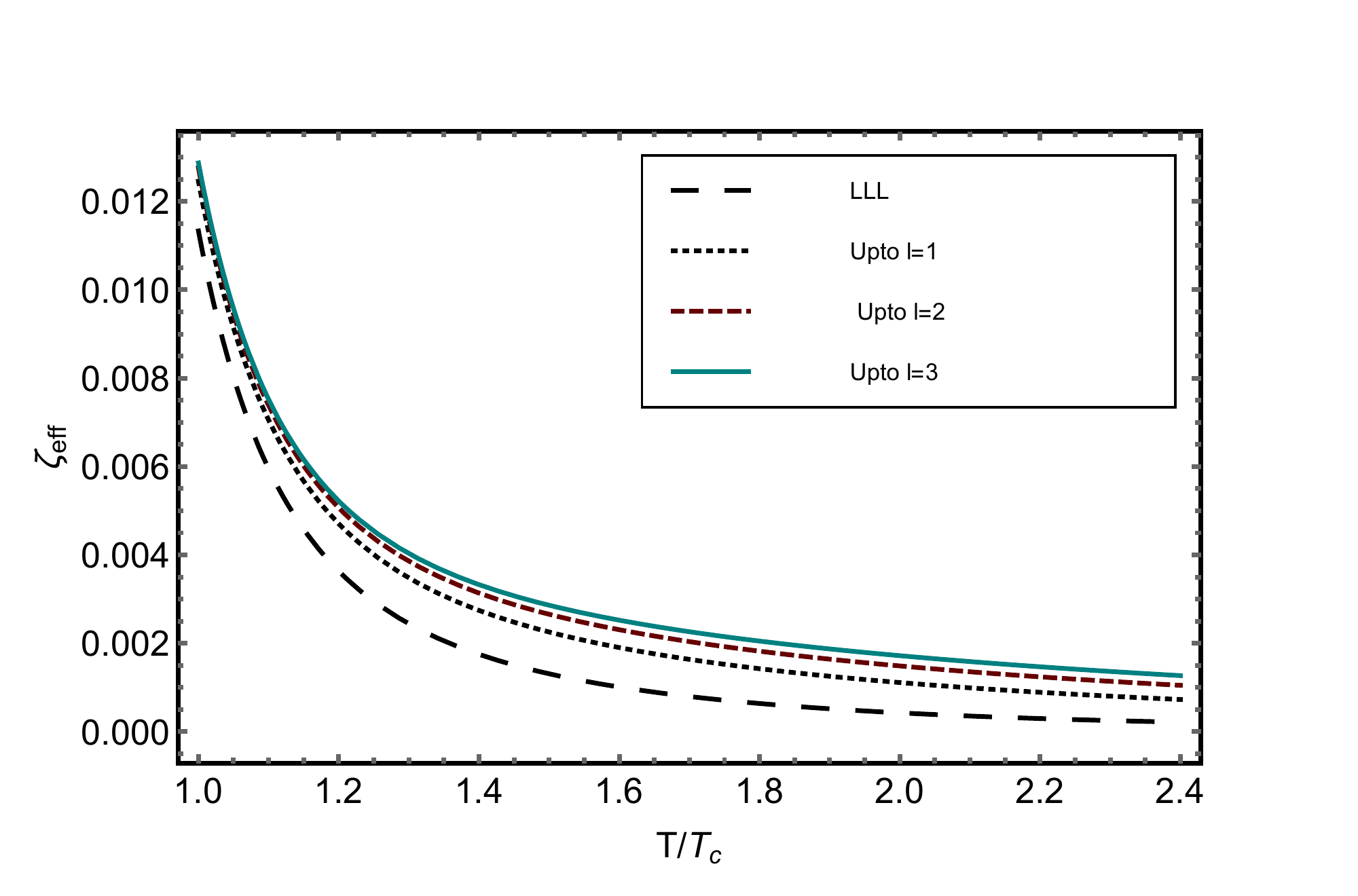}}
\caption{(color online) Effects of HLLs on the bulk viscosity in 
the strong magnetic field $\mid eB\mid=0.3$ GeV$^2$ in the given temperature range.}
\label{f6}
\end{figure}
  
We compared the bulk viscosity to entropy ratio of $1\rightarrow 2$ 
processes with that results from sum 
rule analysis~\cite{Karsch:2007jc} and lattice data 
results~\cite{Meyer:2007dy} as in Fig.~\ref{f5.1}. In~\cite{Karsch:2007jc}, 
the universal properties of  bulk viscosity in the absence 
of magnetic field are studied from the sum rule analysis.
We observe that the magnetic field 
enhances the  $\zeta/s$.
HLLs corrections are significant for the higher 
temperature ranges. We plotted the HLLs corrections to the bulk viscosity in the 
strong magnetic field background in the chosen temperature range 
in Fig.~\ref{f6}. Corrections up to $l=3$ Landau level are shown in the figure.
Higher order corrections beyond third Landau level seems to be negligible in 
the chosen temperature range. Since the HLLs thermal occupation depends 
on $\exp{(-\sqrt{eB}/T)}$, higher order corrections are 
significant at very high temperature. 
 The dominant contributions of the higher order corrections
are entering through the effective coupling and the momentum distribution function. 
Evaluation of the higher order corrections to the matrix element 
of the processes are beyond the scope of this work.

\section{Conclusion and Outlook}
In conclusion, the bulk viscosity of the hot magnetized QCD medium gets 
significant contributions from both the magnetic field and the EoS. The most significant  contributions in the 
strong magnetic field limit to the bulk viscosity  come from the $1\rightarrow 2$ processes in the medium (as these are not possible in the absence of the field).
The bulk viscosity has been computed from semi-classical transport theory approach within the relaxation time approximation.  The thermal relaxation time
for the quarks  is obtained from their respective damping rates in the medium considering the same process. The effects of magnetic fields are encoded in the 
effective quark/antiquark momentum distribution functions in the form of the Landau levels and also in their energy  dispersion relations. On the other hand, the 
gluon dynamics is affected through the effective coupling that has been obtained in our analysis, again following the transport theory approach.

The hot QCD medium effects in the thermal relaxation time of the quarks are found 
to be negligible at very high temperature. Furthermore, the leading order 
term in the bulk viscosity of hot perturbative QCD in strong 
field limit has been estimated from the EQPM using relaxation time approximation and compared against the estimations with and without 
the magnetic field in other approaches. The results in the present work  turned out 
to be consistent with other recent works. All the analysis is done in LLL approximation first, and then 
the effects from the HLLs have been included. The HLLs corrections of 
the bulk viscosity are  found to be quite significant at the higher temperatures. 

We intend to calculate other transport 
coefficients  such as shear viscosity and  charge diffusion coefficient in the strong 
magnetic field background with the EQPM in the near future. Looking at the non-linear aspects of the 
electromagnetic response of the QGP would be another direction to work.

\section*{acknowledgments}
V.C. would like to acknowledge SERB, Govt. of India for the Early
Career Research Award (ECRA/2016). 
We are thankful to Sukanya Mitra for helpful discussions and suggestions. 
We are indebted to the people of India for generous support 
for the research in basic sciences.

\appendix
\section{Boltzmann equation in the linear response regime }\label{A}
We need to solve the Boltzmann equation with appropriate collision 
integral for $1\rightarrow 2$ processes. We have,
\begin{equation}\label{A1}
\left( \partial_t+v_z\partial_z\right)   f^l_q(p_z,t,z)
=-\dfrac{\delta f^l_q}{\tau_{eff}},
\end{equation}
where $\tau_{eff}$ is the thermal relaxation time for $1\rightarrow 2$ process. 
We consider $u_z$ and $\delta f^l_q$ to be small since the 
prime focus is on the linear response regime. 
Using extended EQPM quasiquark momentum distribution defined in 
Eq.~\ref{1}, the Eq.~(\ref{A1}) becomes,
\begin{align}\label{A2}
\delta f^0_{q}=&-\tau_{eff}\bar{f}^0_q(\bar{f}^0_q-1)\nonumber\\
&\times\left[E^0_p  \partial_t\beta+z_q
\partial_tz^{-1}_q-\beta v_zp_z\Theta(z)\right] ,
\end{align}
with $\Theta(z)\equiv(\partial_zu_z)$. Since temperature is time depended, 
Eq.~(\ref{A2}) becomes, 
\begin{align}\label{A3}
\delta f^0_{q}=&-\tau_{eff}\bar{f}^0_q(\bar{f}^0_q-1)\nonumber\\
&\times\left[ (E^0_p -\partial_\beta\ln z_q)
(\partial_t\beta)-\beta v_zp_z\Theta(z)\right].
\end{align}
Finally, we have used $\partial_t\beta=\beta\Omega\Theta(z)$ as defined 
in the Ref~\cite{Hattori:2017qih}. Thus we end up with
\begin{align}\label{A3}
\delta f^0_{q}=&-\tau_{eff}\beta \bar{f}^0_{q}(\bar{f}^0_{q}-1)\Theta(z)\nonumber\\
 &\times[(E^0_p +T^2\partial_T\ln z_q)\Omega-v_zp_z],
\end{align}
where $(E^0_p +T^2\partial_T\ln z_q)\equiv\omega^0_p$ is the single 
particle energy in EQPM.

{}

\end{document}